\documentclass[aps,prx,reprint,superscriptaddress]{revtex4-2}
\usepackage{amsmath} 
\usepackage{xcolor}
\usepackage{graphicx}
\usepackage{physics}
\usepackage{xfrac}
\usepackage{bm}
\usepackage{array, tabularx,  ragged2e,  booktabs}
\usepackage[version=4]{mhchem}
\usepackage{gensymb}
\newcommand{\us}{$\mathrm{\mu}$s}
\usepackage{upgreek}
\usepackage[colorlinks=true]{hyperref}
\hypersetup{linkcolor = [RGB]{206,73,107},
            citecolor = [RGB]{43, 105, 178},
            urlcolor = [RGB]{43, 105, 178}}
            
\begin{document}
\title{Coherent control of optically addressable gallium-vacancy centers in diamond}

\author{Anran Li}
\altaffiliation{These authors contributed equally to this work.}
\author{Sounak Mukherjee}
\altaffiliation{These authors contributed equally to this work.}
\affiliation{Department of Electrical and Computer Engineering, Princeton University, Princeton, NJ 08544, USA}
\author{Rajesh L. Patel}
\author{Matthew L. Markham}
\affiliation{Element Six, Harwell OX11 0QR, United Kingdom}
\author{Nathalie P. de Leon}
\email{npdeleon@princeton.edu}
\affiliation{Department of Electrical and Computer Engineering, Princeton University, Princeton, NJ 08544, USA}

\begin{abstract}
Spin-1 defects with optical addressability and long spin coherence offer attractive platforms for quantum networks and sensing applications. In this work, we report a novel color center containing gallium in diamond that we assign as the negatively charged gallium-vacancy (GaV$^-$) center. We observe optically detected magnetic resonance from an $S=1$ electronic ground state and extract the defect symmetry, zero-field splitting, and nuclear hyperfine constants. We identify three optical zero-phonon lines associated with the defect and propose a tentative electronic level structure from excited-state spectroscopy. Furthermore, we demonstrate coherent control of the spins with favorable coherence properties from cryogenic to room temperature. The defect exhibits microsecond-scale spin relaxation ($T_1$) and spin coherence ($T_2$) times at room temperature. We observe a $T_1$ of 51~ms at 4~K and a $T_2$ of 123~\us{} under dynamical decoupling at 20~K.
\end{abstract}

\maketitle

\section{Introduction}
Optically active spin defects in diamond are leading candidates for solid-state quantum technologies. Among the vast range of color centers in solid-state hosts, defects with spin-1 ground states offer several advantages over spin-\sfrac{1}{2} species \cite{wolfowicz2021quantum, xiong2025identifying}. Their zero-field splitting (ZFS) allows a qubit to be defined and coherently controlled at small magnetic fields.
Some spin-1 defects such as the nitrogen-vacancy (NV) center in diamond \cite{doherty2013nitrogen} can be optically initialized and read out under off-resonant excitation. This enables room-temperature spin control, where phonon-broadened optical transitions hinder spin-selective resonant excitation. Furthermore, a balanced spin configuration suppresses phonon-mediated orbital relaxation in the presence of orbital degeneracy; such orbital relaxation limits spin coherence for group-IV-vacancy centers such as the negatively charged silicon-vacancy center (SiV$^-$) \cite{jahnke2015electron}. SiV$^-$ requires millikelvin temperatures \cite{sukachev2017silicon} or strain engineering to sustain spin coherence at elevated temperatures \cite{klotz2025ultra}. The neutral silicon-vacancy center (SiV$^0$) overcomes this disadvantage, achieving long spin coherence at liquid helium temperatures while preserving a high Debye-Waller factor and inversion symmetry for stable optical transitions \cite{rose2018observation, zhang2020optically}. However, stabilizing the neutral charge state requires careful Fermi level engineering for SiV$^0$ formation \cite{rose2018observation, zhang2023neutral}. Other group-IV defects, including the negatively charged germanium-, tin-, and lead-vacancy centers, have been widely explored for quantum network applications \cite{ruf2021quantum}. Their neutral counterparts \cite{thiering2019eg} have not been experimentally established as optically addressable spin defects thus far \cite{komarovskikh2018spin, krivobok2020observation}. This motivates the investigation of group-III split-vacancy centers in diamond, whose negative charge states are isoelectronic to SiV$^0$, forming $S=1$ ground states with inversion symmetry. They are also predicted to be thermodynamically stable in intrinsic diamond \cite{harris2020group}, simplifying their stabilization and formation.

Here, we report the experimental observation of a gallium-related spin-1 defect in diamond. We observe optically detected magnetic resonance (ODMR) from cryogenic to room temperature. From the magnetic field dependence of ODMR, we extract the defect symmetry, ZFS, and nuclear hyperfine constants. Combining photoluminescence (PL) emission and excitation spectra with the wavelength dependence of optical spin polarization (OSP) and ODMR, we identify the optical transitions and propose a tentative energy-level structure. Based on these measurements, we assign the defect as the negatively charged gallium-vacancy (GaV$^-$) center. Finally, we demonstrate coherent control of the spin defect and characterize its spin relaxation ($T_1$) and coherence ($T_2$) times at different temperatures.

\section{Results}
Electronic-grade single-crystal diamonds (Element Six) are implanted with $^{69}$Ga ions and annealed at 1200\celsius{}, followed by a high-pressure, high-temperature (HPHT) anneal at 1850\celsius{} (Appendix \ref{si:sample}). PL emission is measured with off-resonant excitation at 633, 635, or 642~nm (Appendix \ref{si:plsetup}). The sample temperature is 20~K unless specified otherwise.

\begin{figure}[t!]
\includegraphics{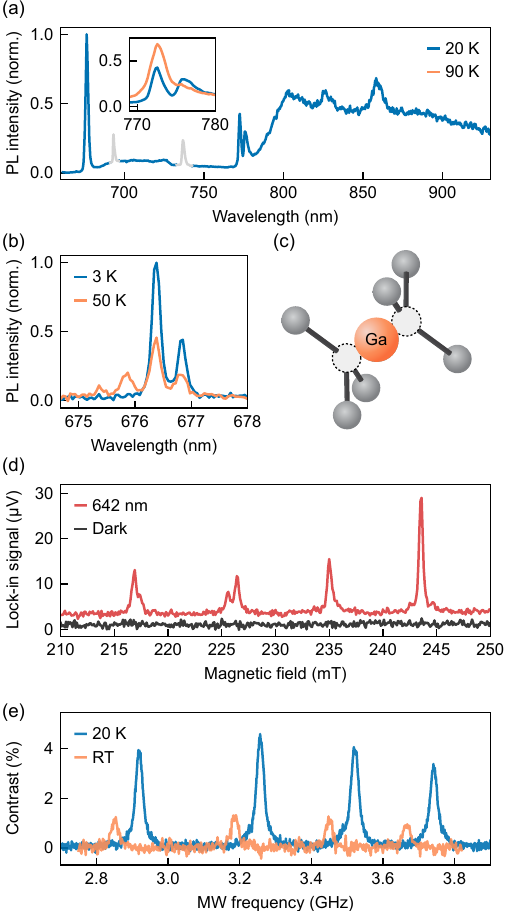} 
    \caption{Photoluminescence (PL) emission and spin resonance. 
    (a) PL spectrum of a high-dose gallium-implanted diamond at 20~K with off-resonant excitation. The Raman line and background emission from SiV$^-$ are grayed out. Inset shows a magnified view of the 773 and 776~nm peaks at 20~K (blue) and 90~K (orange). 
    (b) PL spectrum of a low-implantation-dose sample resolves multiple transitions around 676~nm. 
    (c) Ball-and-stick model of the gallium-vacancy center in diamond. Gray spheres represent carbon atoms and dashed circles represent vacancies.
    (d) Electron spin resonance at 4~K in the presence (red) and absence (black) of optical spin polarization. Four peaks arise from the $^{69}$Ga nuclear hyperfine interaction. The magnetic field is oriented along a $\langle111\rangle$ crystal axis, and the observed splittings in the individual peaks are caused by minor field misalignment. 
    (e) Optically detected magnetic resonance at 20~K (blue) and room temperature (orange).}
    \label{fig1}
\end{figure}

\subsection{Photoluminescence and optical spin polarization}
\label{pl}
Comparing PL from multiple samples under off-resonant excitation, we identify three emission lines at 676.4, 772.6, and 775.8~nm that are unique to gallium-implanted samples. For brevity, we refer to these transitions as 676, 773, and 776~nm, unless precise wavelengths are noted. We observe that HPHT annealing significantly enhances all gallium-related emission (Appendix \ref{si:pl}), as has been demonstrated for other heavy-atom defects in diamond \cite{iwasaki2017tin,abe2026narrow}. Fig.~\ref{fig1}(a) shows the PL spectrum of a high-implantation-dose sample, which is used for all subsequent measurements unless otherwise noted. The 676~nm zero-phonon line (ZPL) is accompanied by a phonon sideband from $\sim$680--765~nm. By integrating the area under the curve and subtracting background emission from the Raman line and SiV$^-$ [Fig.~\ref{fig1}(a)], we estimate a Debye-Waller factor of $\sim$33\% for the 676~nm ZPL. Further to the red, a broad phonon sideband spans $\sim$780--1000~nm, including three characteristic maxima around 800, 825, and 860~nm. Debye-Waller factor estimates for the 773 and 776~nm ZPLs are complicated by their overlapping sidebands. Unlike the 676~nm and 776~nm peaks, the 773~nm peak attains maximum intensity at $\sim$90~K [Fig.~\ref{fig1}(a) inset], which indicates the existence of a lower-lying trapping state (Appendix \ref{si:pl}). With a higher-resolution spectrometer grating, a low-implantation-dose sample reveals fine structure of the 676~nm ZPL [Fig.~\ref{fig1}(b)]. The central peak at 676.4~nm is accompanied by a lower-energy transition at 676.85~nm, emitted from the same excited state to a higher-lying vibronic level in the ground state. From the temperature dependence of PL, we also identify a higher-lying excited state spaced 1.49~meV apart (Appendix \ref{si:pl}). 

Next, we measure the high-implantation-dose sample in a continuous-wave electron spin resonance (CW ESR) spectrometer (Appendix \ref{si:esr}).  Fig.~\ref{fig1}(d) shows the CW ESR spectrum at 4~K with the magnetic field aligned to a $\langle 111 \rangle$ axis of the diamond. We operate in the rapid-passage regime of CW ESR \cite{harbridge2002enhanced}, with the static magnetic field modulated at 100~Hz. This gives rise to absorption peaks instead of the conventional derivative lineshape in phase-sensitive detection \cite{weger1960passage}. We observe four resonance peaks centered at 216.9, 226.4, 235.0, and 243.5~mT under continuous illumination with a 642~nm laser. The spectrum shows the characteristics of an $S=1$ defect with four hyperfine transitions, consistent with the $I=3/2$ nuclear spin of $^{69}$Ga. Additional splittings can be resolved in some of the transitions, attributed to the misalignment of the magnetic field with the crystal axes (Appendix \ref{si:hamiltonian}). In the absence of illumination, the spin polarization drops to thermal equilibrium (5.8\% at 4~K), spins repolarize at the spin relaxation rate rather than the optical pumping rate, and a long $T_1$ fails to restore the population difference within each modulation cycle. A combination of these factors leads to the ESR signal falling below the detection limit in the dark [Fig.~\ref{fig1}(d)].

\begin{figure*}[t]
\centering
\includegraphics{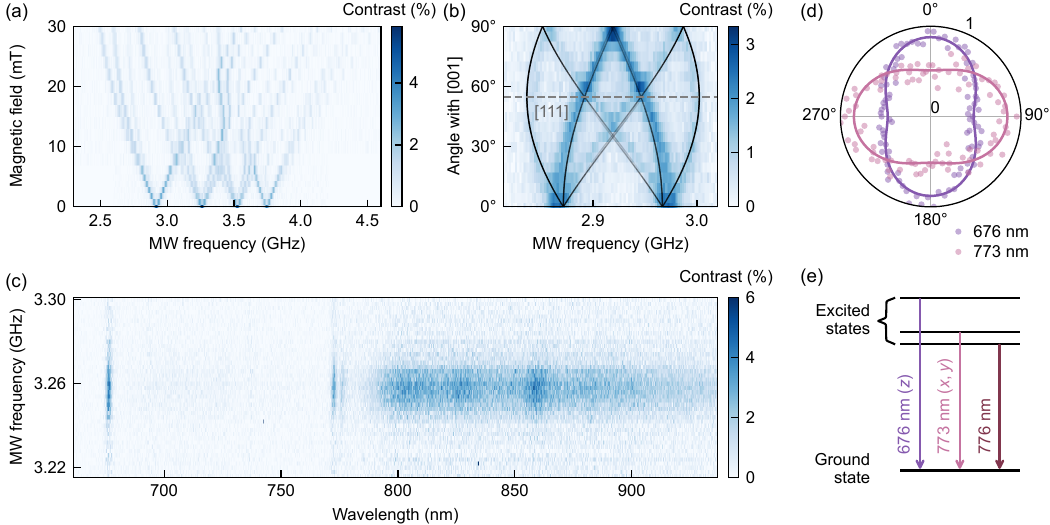} 
    \caption{
    Optically detected magnetic resonance (ODMR). 
    (a) ODMR spectra as a function of magnetic field applied along $[001]$. 
    (b) ODMR spectra as a function of the angle between the [001] axis and the magnetic field. A constant magnitude of 3~mT is used. The dashed line indicates the [111] orientation. 
    (c) Spectrally resolved ODMR at zero field measured at 20~K. The microwave (MW) frequency is scanned across the 3.26~GHz transition, and all three emission lines are enhanced at the same resonance. 
    (d) Normalized ODMR contrast for the 676~nm (purple) and 773~nm (pink) emission lines as a function of the angle of polarization with respect to the defect axis.  Solid lines are sinusoidal fits. A magnetic field of 3~mT is applied to split the ensemble ODMR and isolate a single defect orientation. (e) Tentative electronic structure of the gallium-vacancy center, associated with three emission lines.} 
    \label{fig2}
\end{figure*}

\subsection{Optically detected magnetic resonance}
With OSP established by ESR, we now investigate spin-dependent fluorescence. The spin ensemble is excited with a 635~nm laser, and emission above 650~nm is collected. By measuring PL as a function of the applied microwave frequency, we observe four ODMR peaks with positive contrast at 2.92, 3.26, 3.52, and 3.74~GHz [Fig.~\ref{fig1}(e)]. This is consistent with the ESR spectrum, showing the characteristic $I=3/2$ nuclear hyperfine splitting. At 20~K, a maximum contrast of 20\% is achieved with optimal microwave power delivery under off-resonant excitation [Fig.~\ref{fig:zfs}(a)]. ODMR is observable up to room temperature with the ZFS energy decreasing with temperature (Appendix \ref{si:zfs}).

We characterize the ground-state Hamiltonian by measuring the magnetic field dependence of the ODMR spectrum. Fig.~\ref{fig2}(a) shows that each of the four resonance peaks observed at zero field is split into a group of transitions when an external magnetic field is applied along the [001] direction. To examine the symmetry of the defect, we measure the ODMR spectrum as a function of the magnetic field orientation, rotated in the $(\bar{1}10)$ plane [Fig.~\ref{fig2}(b)]. Here, we focus on the lowermost hyperfine transitions. The angular dependence indicates that the defect has a  $C_3$ axis of symmetry along $\langle 111 \rangle$. The out-of-plane axes are degenerate, establishing a \{110\} mirror plane. This observed pattern restricts the point group of the defect to $D_{3d}$ or $C_{3v}$. Hence, we model the ground-state Hamiltonian with 
\begin{equation}
H = g_e \mu_B \mathbf{B}\cdot\mathbf{S} + \mathbf{S}\hat{\mathbf{D}} \mathbf{S} + \mathbf{S} \hat{\mathbf{A}} \mathbf{I}
\label{hamil}
\end{equation}
where $g_e=2.0023$ is the free-electron $g$-factor, $\mu_B$ is the Bohr magneton, $\mathbf{B}$ is the magnetic field, $\mathbf{S}$ ($\mathbf{I}$) is the electron (nuclear) spin operator vector,
$\hat{\mathbf{D}} = \operatorname{diag}(-D/3, -D/3, 2D/3)$ is the ZFS tensor, and $\hat{\mathbf{A}} = \operatorname{diag}(A_{\perp}, A_{\perp}, A_{\parallel})$ is the hyperfine tensor, both expressed in the defect frame with $\hat{z}\parallel \langle 111 \rangle$. For an $S=1$, $I=3/2$ system, the basis states can be labeled by $|m_S, m_I\rangle$ where $m_S\in\{0, \pm1\}$ and $m_I\in\{\pm 3/2, \pm 1/2\}$. Due to nonzero $A_{\perp}$, the eigenstates at zero field are superpositions of the basis states, leading to uneven spacings between the four transitions (Appendix \ref{si:hamiltonian}). The mixing increases with the off-axis magnetic field, producing the ODMR splittings observed in Fig.~\ref{fig2}(a). Fitting the angular dependence of the $\ket{0,-3/2}\leftrightarrow\ket{-1,-3/2}$ and $\ket{0,+3/2}\leftrightarrow\ket{+1,+3/2}$ transitions to Eq.~\eqref{hamil} yields $D=3.2717\pm0.0014$~GHz, $A_{\parallel}=-260.5\pm0.5$~MHz, and $A_{\perp}=-284.8\pm2.2$~MHz. Details of the Hamiltonian fit are described in Appendix \ref{si:hamiltonian}. The fitted transitions are overlaid on the data in Fig.~\ref{fig2}(b), showing close agreement. The Hamiltonian also explains the magnetic field orientation dependence of the ESR spectrum [Fig.~\ref{fig:esr_orientation}(c)]. We note that the GaV$^-$ center in diamond is predicted to have an $S=1$ ground state with $D_{3d}$ symmetry \cite{harris2020group}. The hyperfine parameters from our fit are in good agreement with theoretical calculations for GaV$^-$ \cite{goss2024density}. Hence, we tentatively assign the gallium-related defect under consideration to GaV$^-$ [Fig.~\ref{fig1}(c)].

To identify the optical transitions associated with the triplet ground state, we perform spectrally resolved ODMR. This is done by scanning the microwave frequency across the resonance at 3.26~GHz and recording the contrast as a function of emission wavelength on a spectrometer. PL associates three emission lines with gallium implantation (Sec.~\ref{pl}). Fig.~\ref{fig2}(c) shows that all three peaks attain maximal contrast at the same resonance frequency, establishing that they originate from the same $S=1$ defect with a ZFS of 3.27~GHz as quantified before. 

Understanding the relationship between these different ZPLs provides crucial insight into the electronic structure of the defect. Since the polarization of an optical transition reflects the symmetry of the states involved, we examine the polarization dependence of each emission line. As all four $\langle 111\rangle$ orientations of the defect contribute to the PL, the ensemble emission does not exhibit polarization dependence. However, by applying a magnetic field, we can isolate the contribution of a single defect orientation and analyze the polarization of spectrally resolved ODMR. Plotting the ODMR contrast as a function of the analyzer angle for two ZPLs [Fig.~\ref{fig2}(d)] reveals that the 676~nm emission is polarized along the [111] axis, whereas the 773~nm emission lies in the orthogonal plane. The 776~nm emission and the phonon sideband show the same polarization as the 773~nm ZPL [Fig.~\ref{fig:wvdep}(a)]. 

\subsection{Excited-state spectroscopy}
\begin{figure}[t]
\includegraphics{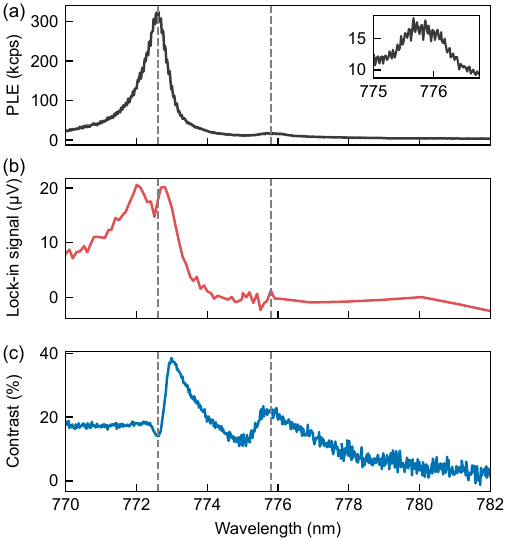}
    \caption{Excited-state spectroscopy. (a) Photoluminescence excitation (PLE) spectrum at 10~K. Excitation power is stabilized at 2~mW. Inset shows a magnified view of the lower-intensity peak at 775.8~nm. (b) CW ESR signal as a function of excitation wavelength measured for the 243.5~mT transition at 4~K. (c) ODMR contrast as a function of excitation wavelength at 10~K. The microwave frequency is fixed at  3.262~GHz. Dashed lines mark the two ZPLs under study, 772.6 and 775.8~nm.}
    \label{fig3}
\end{figure}

To further investigate the ODMR mechanism and excited-state structure, we use a narrow-linewidth ($<$~100~kHz) laser for resonant excitation and combine multiple spectroscopy techniques. Limited by the availability of lasers around 676~nm, we focus our studies on the 773 and 776~nm ZPLs. First, we perform photoluminescence excitation (PLE) spectroscopy, where emission from the phonon sideband above 800~nm is recorded as a function of the excitation wavelength [Fig.~\ref{fig3}(a)]. Two peaks at 772.6~nm and 775.8~nm are observed as expected, with the 775.8~nm peak $\sim$50$\times$ lower in intensity compared to the 772.6~nm peak. This is in sharp contrast to the PL under off-resonant excitation at the same temperature, where the 775.8~nm peak is stronger than 772.6~nm peak. This indicates that the excited state associated with the 775.8~nm ZPL has a lower transition dipole moment and only emits weakly under resonant excitation. With off-resonant excitation, population can undergo nonradiative decay from higher-lying excited states to the 775.8~nm state, leading to observable PL emission. 

Subsequently, we measure the wavelength dependence of OSP using ESR. Fig.~\ref{fig3}(b) shows the ESR signal from the $\ket{0,+3/2}\leftrightarrow\ket{+1,+3/2}$ transition at 243.5~mT as a function of the OSP wavelength. The peak around 773~nm confirms the association of the spin defect observed in ESR with the 773~nm ZPL and transitively to GaV$^-$.  A splitting centered at 772.6~nm is observed, which indicates fine structure of the excited state that we further inspect using ODMR. At 776~nm, the ESR signal falls below the detection limit, suggesting weak OSP. Here, we note that the observed response in CW ESR gives us a qualitative measure of the optical pumping rate and the degree of OSP. Since both can be wavelength dependent, their individual contributions cannot be isolated here. Additionally, in the absence of a thermally polarized reference, we cannot calibrate the ESR signal to quantify the absolute OSP between the $m_S=0$ and $m_S=+1$ spin sublevels.

We then examine the dependence of ODMR contrast on the excitation wavelength [Fig.~\ref{fig3}(c)]. At high excitation energies, the ODMR contrast is comparable to that with off-resonant excitation. The contrast reaches a minimum at 772.6~nm, which corresponds to the PLE maximum. A maximum contrast of 39\% is reached at 773.0~nm. A second peak appears near 775.8~nm, corresponding to the second PLE peak. We note that the wavelength dependence of ODMR contrast is temperature and sample dependent [Figs.~\ref{fig:wvdep}(c) and (d)]. In a low-implantation-dose sample, a higher maximum contrast ($\sim$73\%) is achieved at an excitation wavelength of 772.9~nm.

Detailed spectroscopic characterization with different excitation wavelengths reveals similarities and dissimilarities between SiV$^0$ and GaV$^-$, which shed light on the mechanism of OSP and ODMR. In contrast to our observations of the GaV$^-$ center, SiV$^0$ has limited off-resonant spin polarization, and ODMR is achievable with higher-lying bound exciton states \cite{zhang2020optically}. Furthermore, SiV$^0$ shows wavelength-dependent negative and positive contrast, whereas GaV$^-$ exhibits positive contrast irrespective of the excitation wavelength. Spin-dependent fluorescence of GaV$^-$ under off-resonant excitation is likely to arise from intersystem crossing to a metastable shelving state, analogous to NV centers \cite{doherty2013nitrogen}. Spin-dependent nonradiative decay from the excited state to an intermediate state may also result in a preferred direction of OSP. If a particular spin projection in the excited state manifold gets preferentially shelved, and the population decays back to the same spin sublevel in the ground state, the ``dark'' state with lower fluorescence becomes the optically polarized spin state [Fig.~\ref{fig:mechanism}(a)]. A microwave pulse drives the population from the initial dark state to the bright state, resulting in positive contrast. We note that our measurements cannot specify which one of the $\ket{m_S=0, \pm1}$ sublevels is the dark state. In addition to this mechanism, the observed OSP may also arise from spin-dependent cyclicity of optical transitions [Fig.~\ref{fig:mechanism}(b)]. Optical pumping through this mechanism explains the wavelength dependence of ODMR, which is qualitatively similar to SiV$^0$ \cite{zhang2023neutral}. Although individual optical transitions cannot be resolved within the inhomogeneous linewidth of PLE, resonant excitation around the ZPL reveals the presence of excited-state fine structure in OSP and ODMR.

Now, we combine the experimental evidence from ODMR and excited-state spectroscopy to propose an energy-level structure of the defect [Fig.~\ref{fig2}(e)], based on the tentative assignment to GaV$^-$. We adopt the $D_{3d}$ symmetry predicted by density functional theory (DFT) for a split-vacancy center, which is the energetically favorable configuration for heavy atoms in diamond. \textit{Ab initio} calculations show that the ground-state configuration of GaV$^-$ is $e_u^4 e_g^2$, and its excited-state configuration is $e_u^3 e_g^3$ \cite{harris2020group}. The symmetry of the triplet ground state can be derived to be $^3 A_{2g}$, analogous to SiV$^0$ \cite{green2019electronic, goss2024density}. Promoting an electron from $e_u$ to $e_g$ yields three triplet excited states: $^3E_{u}$, $^3A_{1u}$, and $^3A_{2u}$ \cite{cotton1990chemical}. Transitions from $^3E_{u}$ and $^3A_{1u}$ to $^3 A_{2g}$ are optically allowed, while the transition from $^3A_{2u}$ is dipole forbidden \cite{gali2013ab}. The excited-state configuration is further subject to the $(e_g \otimes e_u) \otimes E_g$ product Jahn-Teller (pJT) effect resulting from JT instabilities of both $e_u$ and $e_g$ orbitals. This effect was first proposed for the neutral group-IV-vacancy centers in diamond \cite{thiering2019eg, ciccarino2020strong}, and extended to the negatively charged group-III split-vacancy defects \cite{harris2020group}. As a result of the pJT effect, the lowest excited states do not retain pure electronic character and instead form vibronic levels, $^3\tilde{E}_u$ and $^3\tilde{A}_{2u}$, where $^3\tilde{A}_{2u}$ has the lowest energy. Thus, we infer the following: (i) The 676.4~nm (1.833~eV) ZPL corresponds to the $z$-polarized $^3A_{1u}$ state. (ii) The 772.6~nm (1.605~eV) ZPL is associated with the $(x,y)$-polarized $^3\tilde{E}_u$ state. (iii) The 775.8~nm (1.598~eV) ZPL with a weak transition dipole moment is linked to $^3\tilde{A}_{2u}$. This transition may be activated by lattice strain induced by high-dose implantation, and it inherits $(x,y)$ polarization from the $^3E_u$ character. Indeed, we find that the 776~nm emission is not observable in the low-implantation-dose sample [Fig.~\ref{fig:samples}(c)]. The thermal occupation of $^3\tilde{E}_u$ over the lowest energy $^3\tilde{A}_{2u}$ state also explains why the 773~nm emission is maximum at an intermediate temperature of $\sim$90~K, similar to the 946~nm ZPL of SiV$^0$ \cite{dhaenens2011optical}. Transition from the $^3A_{1u}$ state of SiV$^0$ has not been experimentally reported \cite{zhang2020optically}. 

\subsection{Spin coherence and relaxation}
\begin{figure}[t]
\includegraphics{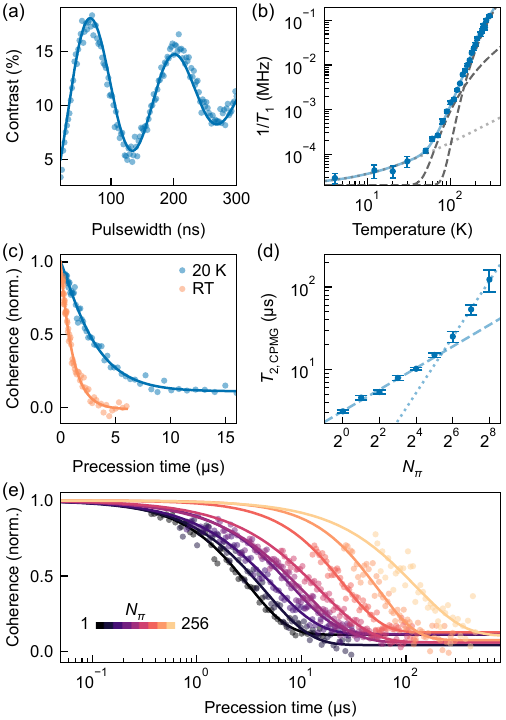} 
    \caption{Coherent control and spin dynamics. 
    (a) Rabi oscillations measured at 20~K. The data points are fitted to $ae^{-t/\tau} \cos(\Omega t + b) + c$ with a decay constant, $\tau=211\pm13$~ns, and Rabi frequency, $\Omega/2\pi = 7.44\pm0.04$~MHz. (b) Temperature dependence of spin relaxation rate, $1/T_1$. The solid line is a fit to the data with Eq.~\eqref{T1rate}. The dotted and dashed gray lines correspond to the direct and Orbach components of the fitted curve, respectively. (c) Spin coherence time ($T_2$) measured at 20~K (blue) and room temperature (orange) with the Hahn echo sequence. The decay is fitted to a stretched exponential, $e^{-(t/T_2)^n}$, with $T_2=3.12\pm0.13$~\us{}, $n=1.14\pm0.09$ at 20~K and $T_2=1.09\pm0.05$~\us{}, $n=1.00\pm0.08$ at room temperature.
    (d) $T_{2,\text{CPMG}}$ as a function of the number of $\pi$-pulses, $N_{\pi}$, at 20~K. The dashed line is a fit to the data for $N_{\pi}$ from 1 to 32, yielding a scaling power of $0.43\pm0.02$. The dotted line is a fit to the data for $N_{\pi}$ from 64 to 256, yielding a scaling power of $1.13\pm0.21$. (e) Coherence under dynamical decoupling using the CPMG sequence with $N_{\pi}$=1, 2, 4, ..., 256 (dark to light). All coherence measurements are performed on the 2.92~GHz hyperfine transition.}
    \label{fig4}
\end{figure}

With an understanding of the optical initialization and readout from ODMR, we characterize the spin dynamics of GaV$^-$. First, we realize coherent control of the spins using pulsed ODMR on the lowermost hyperfine transition and observe Rabi oscillations [Fig.~\ref{fig4}(a)]. Next, we measure spin relaxation times of GaV$^-$ across a range of temperatures: $T_1=50.8\pm13.3$~ms at 4~K and $T_1=7.85\pm0.50$~\us{} at room temperature. The temperature dependence of the spin relaxation rate is shown in Fig.~\ref{fig4}(b). We observe that the relaxation rate increases linearly until $\sim$40~K, followed by an exponential rise at elevated temperatures. This is consistent with the picture that $T_1$ is governed by a direct process via single phonons at low temperatures and Orbach processes at high temperatures. Hence, we model the relaxation rate with \cite{mukherjee2026defect}
\begin{equation}
\frac{1}{T_1} = \frac{1}{T_{1_\text{sat}}} + \Gamma_{\text{direct}} T +  \frac{\Gamma_{\text{Orbach}_1}}{e^{E_1/k_BT}-1} + \frac{\Gamma_{\text{Orbach}_2}}{e^{E_2/k_BT}-1}.
\label{T1rate}
\end{equation}
Fitting the data yields $T_{1_\text{sat}}=49.95\pm6.67$~ms, Orbach activation energies $E_1=32.67\pm3.38$~meV and $E_2=86.66\pm2.92$~meV, coupling rates $\Gamma_{\text{Orbach}_1}=40.19\pm16.02$~kHz, $\Gamma_{\text{Orbach}_2}=3.24\pm0.40$~MHz, and $\Gamma_{\text{direct}}=1.60\pm0.42$~Hz/K. Similar trends have been reported for NV centers and the WAR5 defect in diamond, where the Orbach modes are attributed to vibrational modes of the impurity atoms and carbon atoms \cite{cambria2023temperature, mukherjee2026defect}.

Finally, we measure the spin coherence using the Hahn echo sequence: $T_2=3.12\pm0.13$~\us{} at 20~K and $T_2=1.09\pm0.05$~\us{} at room temperature [Fig.~\ref{fig4}(c)]. $T_2$ is significantly shorter than the millisecond-scale $T_1$ at cryogenic temperatures. To extend the coherence and study the noise bath, we use dynamical decoupling. $T_2$ measured with the Carr-Purcell-Meiboom-Gill (CPMG) sequence for a varying number of $\pi$-pulses, $N_{\pi}$, is shown in Fig.~\ref{fig4}(d). With 256 $\pi$-pulses, $T_{2,\text{CPMG}}=123.33\pm36.26$~\us{} [Fig.~\ref{fig4}(e)]. Using CPMG as a filter function, we can compute the noise spectrum $S(\omega)$ of the spin bath \cite{BarGill2012}, shown in Fig.~\ref{fig:SD}(a). We find that the best fit to $S(\omega)$ is a double Lorentzian model, with fitted correlation times $\tau_{c_1}=1.05\pm0.10$~\us{} and $\tau_{c_2}=102\pm44$~ns (Appendix \ref{si:sd}). Since the observed $T_2$ is shorter than typical coherence times for NV centers in electronic-grade diamonds, we exclude substitutional nitrogen ($<5$~ppb) and $^{13}$C nuclear spins (1.1\% natural abundance) as dominant noise sources \cite{bauch2020decoherence}. Instead, the two Lorentzian noise baths may originate from irradiation-induced defects, gallium nuclear spins, or unidentified gallium-related electron spins. We note that the dependence of $T_{2,\text{CPMG}}$ on $N_{\pi}$ deviates from a single power law, $N_\pi^\gamma$, which can be explained by the noise spectrum. For $1 \le N_\pi \le 32$, we find the scaling power $\gamma = 0.43 \pm 0.02$, while for $64 \le N_\pi \le 256$, $\gamma = 1.13 \pm 0.21$ [Fig.~\ref{fig4}(d)]. 
At smaller $\omega$, the superposition of two Lorentzian lineshapes yields $S(\omega) \propto \omega^{-1}$. At larger $\omega$, the shape of $S(\omega)$ becomes dominated by the second Lorentzian bath with the shorter correlation time, approaching the $\omega^{-2}$ scaling. The change of the shape of $S(\omega)$ hence gives rise to the different CPMG scaling powers at different $N_{\pi}$, equivalent to different frequency regimes being probed.

\section{Summary and Outlook}
In conclusion, we report the experimental observation of a novel spin defect containing gallium in diamond. The defect exhibits an $S=1$ ground state with high ODMR contrast from liquid helium to room temperature. We identify three ZPLs of the defect and propose an energy-level structure, which we assign to GaV$^-$. We demonstrate coherent control of the spin ensemble and observe microsecond-scale $T_1$ and $T_2$ at room temperature. At 20~K, the spin coherence is limited by the noise bath consisting of implantation-induced defects, and it can be extended to more than 100~\us{} using dynamical decoupling. We point out that a substitutional position of the impurity atom with NV-like $C_{3v}$ symmetry gives rise to only one triplet excited state \cite{doherty2013nitrogen}. The selection rules observed for the three excited-state transitions support our assignment to the GaV$^-$ split-vacancy center with $D_{3d}$ symmetry. A complete theoretical description of the excited-state structure, along with accurate ZPL calculations taking into account the spin-orbit coupling, can further verify our assignment and remains an open task. Future work includes detailed characterization of the spin coherence and optical properties of single centers; preliminary measurements of low-implantation-dose samples reveal blinking and spectrometer-limited optical linewidths (Appendix \ref{si:single}). In particular, the putative non-polar symmetry of the defect should give rise to enhanced spectral stability compared to NV centers, making GaV$^-$ a potential candidate quantum memory and spin-photon interface for quantum networks.

The favorable spin properties of GaV$^-$ at elevated temperatures also make it attractive for quantum sensing. NV centers suffer from PL quenching at intermediate temperatures \cite{ernst2023temperature, happacher2023temperature}, where the gap can be filled in by defects with a temperature-robust optical readout. From continuous-wave ODMR, we estimate DC magnetic-field sensitivities of 6.37~$\mu$T~Hz$^{-\sfrac{1}{2}}$ at 10~K and 17.46~$\mu$T~Hz$^{-\sfrac{1}{2}}$ at room temperature, which can be improved by achieving narrower spin linewidths. A systematic study of post-implantation annealing conditions could help mitigate strain, achieve higher conversion yields, and improve inhomogeneous linewidths \cite{iwasaki2017tin, gorlitz2020spectroscopic}. More broadly, the experimental realization and robust formation of gallium-vacancy centers in diamond motivate the search for other group-III split-vacancy centers as alternative platforms for quantum applications.

\textit{Note added.}-- During the preparation of this manuscript, we became aware of an independent work on the ODMR and spin coherence of gallium-vacancy centers in diamond \cite{dodson2026optically}.

\begin{acknowledgments}
We gratefully acknowledge Alex Abulnaga, Sean Karg, and Zi-Huai Zhang for help with sample preparation and Zhiyang Yuan for help with PCB design. We also thank Yilin Chen and Chris G. Van de Walle for fruitful discussions. This work was primarily supported by the Center for Molecular Quantum Transduction (CMQT), an Energy Frontier Research Center funded by the U.S. Department of Energy, Office of Science, Basic Energy Sciences, under Contract No. DE-SC0021314. The ESR spectrometer and magnetic-field-dependent ODMR (A.L., S.M., and N.P.d.L.) were also supported by the U.S. Department of Energy, Office of Science, National Quantum Information Science Research Centers, Co-design Center for Quantum Advantage (C2QA) under Contract No. DE-SC0012704 and the Eric and Wendy Schmidt Transformative Technology Fund. 
\end{acknowledgments}

\appendix
\section{Experimental details}
\label{si:exp}
\subsection{Sample preparation}
\label{si:sample}
\begin{table}[b]
\renewcommand{\arraystretch}{1.15}
\caption{Recipes used for the high-dose (Sample A) and low-dose (Sample B) $^{69}$Ga implantation.}
\label{tab:fluence}
\begin{ruledtabular}
\begin{tabular}{ccc}
 & \multicolumn{2}{c}{Fluence (ions/cm$^2$)} \\
\cline{2-3}
Energy (keV) & Sample A & Sample B \\
\hline
40    & $3.69 \times 10^{12}$ & $2.22 \times 10^{10}$ \\
80    & $3.69 \times 10^{12}$ & $2.22 \times 10^{10}$ \\
120   & $5.14 \times 10^{12}$ & $3.09 \times 10^{10}$ \\
180   & $5.79 \times 10^{12}$ & $3.48 \times 10^{10}$ \\
240   & $6.94 \times 10^{12}$ & $4.17 \times 10^{10}$ \\
310   & $8.94 \times 10^{12}$ & $5.37 \times 10^{10}$ \\
400   & $1.58 \times 10^{13}$ & $9.50 \times 10^{10}$ \\
\hline
Total & $5 \times 10^{13}$    & $3 \times 10^{11}$ \\
\end{tabular}
\end{ruledtabular}
\end{table}

Electronic-grade diamonds (Element Six) with \{001\} faces are prepared for ion implantation by polishing, reactive ion etching, and surface cleaning, as described in Ref.~\cite{sangtawesin2019origins}. The samples are then implanted with $^{69}$Ga at 0$^\circ$ tilt at a commercial foundry. The energy and fluences for high-dose (Sample A) and low-dose (Sample B) implantation  are specified in Table \ref{tab:fluence}. For single-center measurements (Appendix \ref{si:single}), Sample C is implanted with $10^9$~ions/cm$^2$ at 70~keV. Next, the samples are cleaned and annealed at 1200\celsius{} at low pressures $<3\times10^{-4}$~Pa \cite{sangtawesin2019origins}. Subsequently, they are annealed at Element Six at 1850\celsius{} under a pressure of 8~GPa for 10~minutes. For comparing PL, a boron-doped diamond (Sample D) is implanted with $10^{13}$~ions/cm$^2$ at 105~keV and annealed at 1200\celsius{} only. Using Fourier-transform infrared spectroscopy \cite{howell2019ftir, liggins2009ftir}, we quantify $[\mathrm{B}]\sim10^{17}$~cm$^{-3}$ and $[\mathrm{N_s^+}]\sim1.5\times10^{17}$~cm$^{-3}$ in this sample. SiV$^-$ is observed in all samples because of silicon contamination.

\subsection{PL and ODMR spectroscopy}
\label{si:plsetup}
PL and spin dynamics are measured in a home-built confocal microscope optimized for ensemble PL detection. The sample is mounted on a copper cold finger inside a helium flow cryostat (Janis ST-100). The excitation is provided by an intensity-modulated 633~nm laser (Cobolt 06-MLD 633), filtered by a bandpass filter (Thorlabs FLH635-10). Excitation is focused on the sample with a 10$\times$ near-infrared objective (Olympus LMPLN10XIR) outside the cryostat. Excitation and detection channels are separated by a long-pass dichroic mirror (Thorlabs DMLP650). Emission is filtered with a long-pass filter (Thorlabs FELH0650 or FELH0750) and routed to a single-photon detector (Excelitas Technologies SPCM-900-44-FC) via a single-mode fiber or to a spectrometer (Princeton Instruments SpectraPro HRS-300, PIXIS 100 camera) via a 50~$\mu\mathrm{m}$ core multimode fiber. A grating with 300~grooves/mm is used for broadband spectra, and 1200~grooves/mm for fine resolution. For optical lifetime measurements, a supercontinuum white laser (NKT Photonics SuperK FIANIUM FIR-20) with a bandpass filter (Semrock TBP01-628/14-25x36) centered around 635~nm is used at a repetition rate of 78~MHz. Individual ZPLs are isolated with angle-tuned filters (Semrock FF01-679/41-25, FF01-781/23-25, and FF01-776/LP-25). Photon counts are detected with a time-correlated single photon counter (Swabian Time Tagger Ultra). 

PLE and magnetic field-dependent ODMR measurements are performed in a home-built confocal microscope equipped with a closed-cycle cryostat and a three-axis superconducting magnet (Cryomagnetics C-Mag Vari-611). We use a 635~nm laser (New Focus TLB-6704) for off-resonant excitation and a continuous-wave Ti:sapphire laser (MSquared SolsTiS) for PLE. The resonant excitation wavelength is monitored on a wavemeter (Bristol 721). The sample is mounted onto nanopositioners and piezoelectric scanners (Attocube ANPx311, ANPz102, and ANSxy100) for translation, focusing, and confocal scanning. The excitation beam is focused onto the sample with a high-numerical-aperture apochromatic objective (Attocube LT-APO/NIR/0.81) and PL is detected with a superconducting nanowire single-photon detector (SNSPD, Quantum Opus). Off-resonant excitation and detection paths are similar to the ensemble setup. For resonant excitation, a different dichroic mirror (Semrock FF801-Di02-25x36) and filters (Thorlabs FESH0800 and FELH0800) are used. For polarization-dependent measurements, a rotating half-wave plate (Thorlabs AHWP10M-980) followed by a polarizer (Thorlabs LPVIS050-MP2) is used as an analyzer in the detection path.

For ODMR, microwaves (MW) are delivered via an antenna on a printed circuit board (PCB) attached to the sample mount or objective holder, leaving approximately 50~$\mu$m between the antenna and the sample surface. The MW excitation is generated with a signal generator (Agilent N9310A or Rohde \& Schwarz SMB100A). For ODMR across a broad frequency range, a two-port PCB is used with a wideband amplifier (Mini-Circuits ZVA-183G-S+). For efficient power delivery at the lowermost transition frequency, one of the ports of the PCB is left open and the trace length is optimized to form standing waves \cite{yuan2024instructional}. This is combined with a high-power amplifier (Ophir 5144). An isolator (Pasternack PE83IR1017) is added after the amplifier for circuit protection. The MW excitation is pulsed using a fast MW switch (Mini-Circuits ZASWA-2-50DR+). The timing of MW and optical pulses is synchronized using a programmable pulse generator (SpinCore PBESR-PRO-500), and phase control of the MW pulses is achieved with an arbitrary waveform generator (Keysight 33622A).

\begin{figure}[b!]
\includegraphics{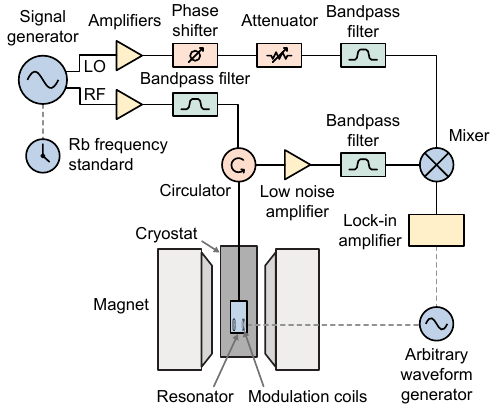}
    \caption{Schematic diagram of the home-built CW ESR setup.}
    \label{fig:esr}
\end{figure}

\subsection{Electron spin resonance}
\label{si:esr}
 CW ESR measurements are performed in a home-built spectrometer operating in the X-band ($\sim$9.7~GHz). Cryogenic temperatures are achieved with a helium flow cryostat (ColdEdge Stinger). The sample is mounted on a sapphire substrate attached to the open end of a Suprasil tube (Wilmad 722-PQ-7) with varnish (VGE-7031), placed inside a sapphire dielectric ring resonator with modulation coils (Bruker ER 4118 X-MD-5-W1). Spins are optically polarized with a 642~nm laser diode (Thorlabs L638P200) via a multimode fiber (Thorlabs M119L03, 400~$\mu$m core diameter) in contact with the edge of the sample. The magnet (Bruker B-E 25, ER083 power supply) is controlled with a field controller and teslameter (Lake Shore F41) equipped with a Hall probe (Lake Shore FP-NS-180-TS05M).

The microwave bridge is a modified version of the pulsed ESR setup described in Ref.~\cite{mukherjee2026defect}, as shown in Fig.~\ref{fig:esr}. Microwave excitation  is provided by a vector signal generator (Agilent E8267D), followed by a high-power amplifier (Pasternack PE15A5071) and a bandpass filter (K\&L 5C50-9625/T250-O/O). A circulator (Pasternack PE83CR1008) separates the excitation and detection paths. On the detection arm, the signal is amplified by a low-noise amplifier (Pasternack PE15A1025) followed by a bandpass filter (Mini-Circuits ZVBP-9750-S+) and down-converted using a mixer (Mini-Circuits ZMX-10G+). For phase-sensitive detection, the static magnetic field is modulated by an arbitrary waveform generator (Keysight 33622A) at 100~Hz with an amplitude of 0.17~mT. The down-converted signal is demodulated by a lock-in amplifier (Stanford Research Systems SR860) referenced to this modulation frequency.

\section{Supplementary PL measurements}
\label{si:pl}
\begin{figure}[b!]
\includegraphics{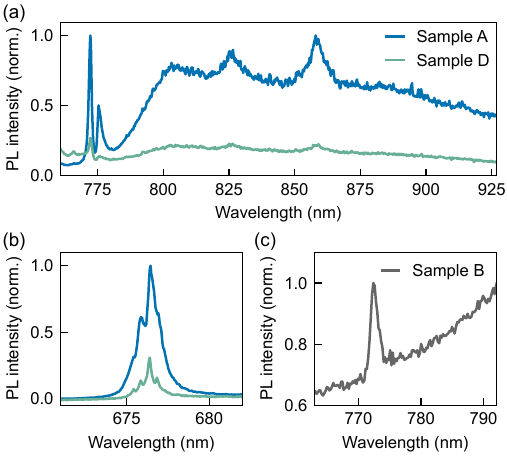}
    \caption{PL spectrum of Sample A (HPHT annealed) compared with Sample D (1200\celsius{} annealed): (a) 773 and 776~nm peaks. (b) Emission lines around 676~nm. PL from Sample D has been scaled up $5\times$ to account for the lower implantation fluence. (c) PL spectrum of Sample B showing the 773~nm ZPL. 776~nm emission is not observed in this sample.}
    \label{fig:samples}
\end{figure}

To find the emission lines originating from gallium, we measure PL across multiple samples. All gallium-implanted samples show emission around 676~nm. The individual transitions are broadened and overlap in Samples A and D [Fig.~\ref{fig:samples}(b)], whereas they are resolvable in the low-dose implantation in Sample B [Fig.~\ref{fig1}(b)]. Samples A and D show both the peaks at 773 and 776~nm [Fig.~\ref{fig:samples}(a)]. 776~nm is not observable in the low-implantation-dose sample [Fig.~\ref{fig:samples}(c)]. This indicates that strain from the higher implantation fluence may activate the 776~nm transition. Comparing the spectra also reveals that HPHT annealing significantly enhances all three emission lines in Sample A, compared to Sample D, which has undergone 1200\celsius{} anneal only. We note that the increase in conversion yield from HPHT anneal cannot be estimated quantitatively from these two samples, as the presence of boron also plays a role in the yield for a particular charge state of defects.

\begin{figure}[t]
\includegraphics[width=1\linewidth]{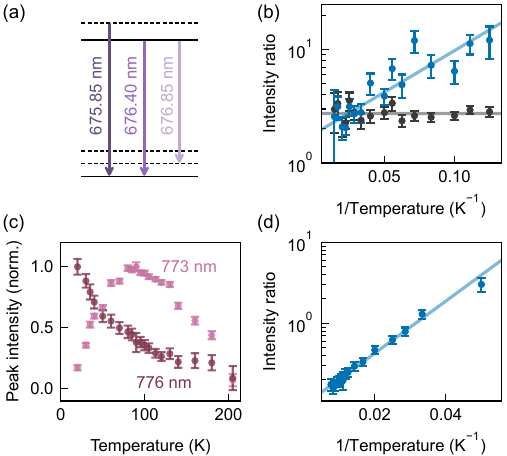} 
    \caption{(a) Tentative level structure of the 676.4~nm ZPL. (b) Ratio of the 676.4~nm peak intensity to the 675.85~nm peak intensity (blue) as a function of the inverse of temperature. The slope yields an activation energy of $1.45\pm0.05$~meV. The black data points show that the intensity ratio for the 676.4 and 676.85~nm peaks is constant across temperatures. (c) Temperature dependence of 773 and 776~nm peak intensities. (d) Ratio of the 776~nm peak intensity to the 773~nm peak intensity, fit to an activation energy of $6.39\pm0.29$~meV.}
    \label{fig:temp}
\end{figure}

\begin{figure}[t]
\includegraphics[width=1\linewidth]{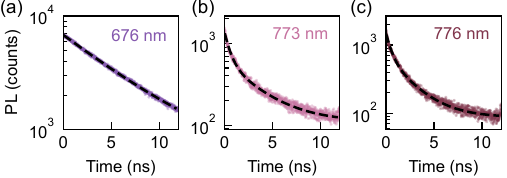} 
    \caption{Time-resolved PL at 20~K for (a) 676~nm, (b) 773~nm, and (c) 776~nm. Dashed black lines are biexponential fits, where the shorter timescale of 0.5~ns is attributed to the instrument response function. The excited-state lifetimes for the 676, 773, and 776~nm transitions are obtained to be $6.54\pm0.04$~ns, $3.37\pm0.09$~ns, and $2.46\pm0.05$~ns, respectively.}
    \label{fig:lifetime}
\end{figure}

We propose a tentative level structure of the transitions around 676~nm from temperature dependence. We observe that the 676.4 and 676.85~nm transitions decrease in intensity monotonically with temperature, whereas the 675.35 and 675.85~nm lines are thermally activated. The ratio of intensities between the 676.4 and 675.85~nm peaks can be fitted to an activation energy of $1.45\pm0.05$~meV [Fig.~\ref{fig:temp}(b)], consistent with their energy spacing (1.49~meV). This points to a higher-lying excited state as the origin of the 675.85~nm emission. On the other hand, the ratio of the 676.4 and 676.85~nm peak intensities is constant across temperatures, indicating that they arise from the same excited state. Peaks at lower energy than the ZPL of 676.4~nm at a spacing of 1.2~meV [Fig.~\ref{fig:single}(d)] may arise from vibronic replicas in the ground state. The origin of the 675.35~nm line is unknown. Similarly, the 773 and 776~nm ZPLs arise from closely spaced excited states separated by 6.6~meV. Thus, at lower temperatures, the population gets trapped in the lower-energy state, quenching the 773~nm PL. This is demonstrated in Fig.~\ref{fig:temp}(c) and the ratio of peak intensities gives us an activation energy of $6.39\pm0.29$~meV. We also measure the excited-state lifetimes of the three ZPLs, as shown in Fig.~\ref{fig:lifetime}.

\section{Hamiltonian calculation and fitting}
\label{si:hamiltonian}
For a spin-1 defect with $C_3$ symmetry, the axial ZFS and hyperfine tensors are described by \cite{cann2009magnetic}
\begin{equation}
\hat{\mathbf{D}} =    
\begin{pmatrix}
-D/3 & 0 & 0\\
0 & -D/3 & 0\\
0 & 0 & 2D/3
\end{pmatrix},\
\hat{\mathbf{A}} =    
\begin{pmatrix}
A_{\perp} & 0 & 0\\
0 & A_{\perp} & 0\\
0 & 0 & A_{\parallel}
\end{pmatrix}
\end{equation}
respectively. For each defect axis ([$111$], [$1\bar{1}\bar{1}$], [$\bar{1}1\bar{1}$], [$\bar{1}\bar{1}1$]), a defect frame $(\hat{n}_x, \hat{n}_y, \hat{n}_z)$ is defined such that $\hat{n}_z$ aligns with the defect axis. The Hamiltonian [Eq.~\eqref{hamil}] is defined in the defect frame, so that under a magnetic field, $\hat{\mathbf{D}}$ and $\hat{\mathbf{A}}$ remain diagonal, and the magnetic field direction defined in the lab frame $\mathbf{B_\text{lab}}$ is transformed to the defect frame by $B_{\text{defect}}^i = \mathbf{B_\text{lab}} \cdot \hat{n}_i$. 

\begin{figure}
\centering
\includegraphics{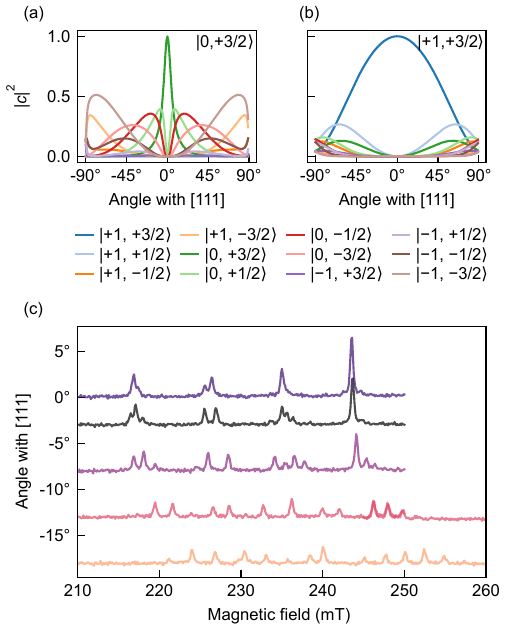}
\caption{Spin mixing with a misaligned magnetic field. Decomposition of the (a) $\ket{0, +3/2}$ and (b) $\ket{+1, +3/2}$ eigenstates into basis states as a function of the angle between [111] and the magnetic field, calculated from the fitted Hamiltonian. The magnitude of the magnetic field is fixed at 243.5~mT to model the behavior in the ESR regime. (c) ESR as a function of the angle between [111] and the magnetic field.}
\label{fig:esr_orientation}
\end{figure}

\begin{figure}
\centering
\includegraphics{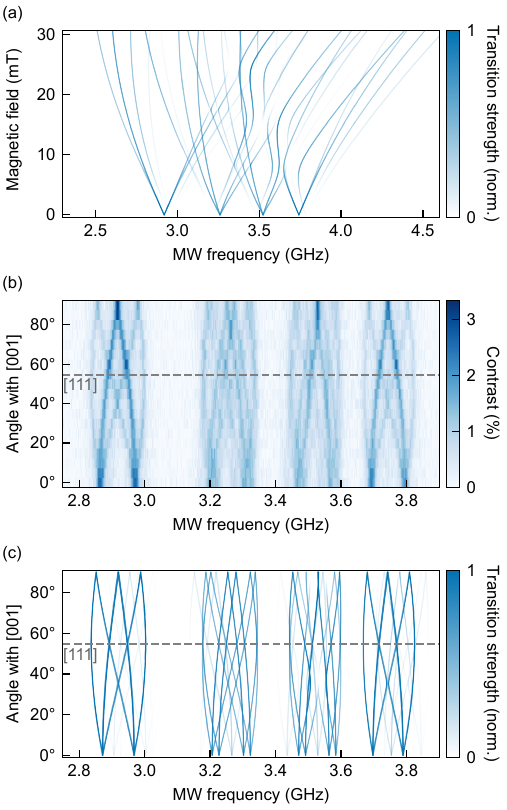}
\caption{ODMR spectra as a function of the magnetic field magnitude and orientation. (a) Calculated transition frequencies for varying magnetic field along $[001]$, for the experimental conditions in Fig.~\ref{fig2}(a). (b) ODMR spectrum as a function of the angle between the $[001]$ axis and the magnetic field. The magnetic field is kept constant at 3~mT. (c) Calculated transition frequencies for the conditions in (b) obtained from the fitted Hamiltonian. Gray dashed lines indicate the $[111]$ direction.}
\label{fig:hamiltonian_fit}
\end{figure}

There are 12 basis states in total, labeled $|m_S, m_I\rangle$ where $m_S\in\{0, \pm1\}$ and $m_I\in\{\pm3/2, \pm1/2\}$.
Spin mixing occurs as a result of the transverse hyperfine interaction $A_{\perp}$. Table~\ref{tab:eigenstates} shows the calculated eigenstates at zero field, written as $\psi = \sum_i c_i |m_S, m_I\rangle_i$, where each entry lists $c_i$ of the basis state $i$. For higher fields aligned with the defect axis, where the hyperfine interaction is small compared to the Zeeman splitting, spin mixing can be neglected. The mixing behavior increases with misaligned magnetic fields. As an example, the cases for two eigenstates, $\ket{0,+3/2}$ and $\ket{+1,+3/2}$, at the ESR field of 243.5~mT are shown in Figs.~\ref{fig:esr_orientation}(a) and (b), respectively. Hence, previously forbidden transitions with $\Delta m_I \neq0$ are enabled at misaligned magnetic fields, giving rise to additional resonance features in the ESR [Fig.~\ref{fig:esr_orientation}(c)] and ODMR spectra [Fig.~\ref{fig:hamiltonian_fit}]. To fit the Hamiltonian, each transition needs to be assigned to the defect axis and the exact pair of eigenstates. To simplify the fitting process, we use the data from Fig.~\ref{fig2}(b) along with the zero-field ODMR transitions as the dataset for fitting. For each parameter vector $\mathbf{p} = (D, A_\parallel, A_\perp)$, we define
\begin{equation}
  \chi^2(\mathbf{p}) = \sum_{k}
\left[\frac{f_k(\mathbf{p};\mathbf{B}_k) - f_k^{\mathrm{exp}}}{\sigma_k}\right]^2,
\label{eq:chi2}
\end{equation}
where $f_k^{\mathrm{exp}}$ is the measured frequency at field $\mathbf{B}_k$ and $\sigma_k$ is its uncertainty. $\chi^2$ is minimized using a least-squares method. All Hamiltonian calculations are performed with the QuTiP package in Python \cite{qutip5}. The complete ODMR spectra and calculations from the fitted Hamiltonian are shown in Fig.~\ref{fig:hamiltonian_fit}. The remaining discrepancies between the data and the calculations could originate from misalignment and calibration errors of magnetic fields. 

\begin{table*}\caption{Calculated eigenstates at zero field. }
\label{tab:eigenstates}
\begin{ruledtabular}
\begin{tabular}{l*{12}{c}}
 & \multicolumn{4}{c}{$m_S=+1$} & \multicolumn{4}{c}{$m_S=0$} & \multicolumn{4}{c}{$m_S=-1$} \\
\cmidrule(lr){2-5}\cmidrule(lr){6-9}\cmidrule(lr){10-13}
Energy (GHz)
 & $+3/2$ & $+1/2$ & $-1/2$ & $-3/2$
 & $+3/2$ & $+1/2$ & $-1/2$ & $-3/2$
 & $+3/2$ & $+1/2$ & $-1/2$ & $-3/2$ \\
\hline
-2.260 & 0 & 0 & 0.114 & 0 & 0 & 0.989 & 0 & 0 & 0.092 & 0 & 0 & 0 \\
-2.260 & 0 & 0 & 0 & 0.092 & 0 & 0 & 0.989 & 0 & 0 & 0.114 & 0 & 0 \\
-2.219 & 0 & 0 & 0 & 0 & 0 & 0 & 0 & 0.994 & 0 & 0 & 0.109 & 0 \\
-2.219 & 0 & 0.109 & 0 & 0 & 0.994 & 0 & 0 & 0 & 0 & 0 & 0 & 0 \\
0.700 & 1 & 0 & 0 & 0 & 0 & 0 & 0 & 0 & 0 & 0 & 0 & 0 \\
0.700 & 0 & 0 & 0 & 0 & 0 & 0 & 0 & 0 & 0 & 0 & 0 & 1 \\
0.999 & 0 & 0.994 & 0 & 0 & -0.109 & 0 & 0 & 0 & 0 & 0 & 0 & 0 \\
0.999 & 0 & 0 & 0 & 0 & 0 & 0 & 0 & -0.109 & 0 & 0 & 0.994 & 0 \\
1.261 & 0 & 0 & 0 & -0.157 & 0 & 0 & -0.099 & 0 & 0 & 0.983 & 0 & 0 \\
1.261 & 0 & 0 & 0.983 & 0 & 0 & -0.099 & 0 & 0 & -0.157 & 0 & 0 & 0 \\
1.520 & 0 & 0 & 0.146 & 0 & 0 & -0.109 & 0 & 0 & 0.983 & 0 & 0 & 0 \\
1.520 & 0 & 0 & 0 & 0.983 & 0 & 0 & -0.109 & 0 & 0 & 0.146 & 0 & 0 \\
\end{tabular}
\end{ruledtabular}
\end{table*}

\section{Temperature dependence of ZFS}
\label{si:zfs}
\begin{figure}[t]
\includegraphics{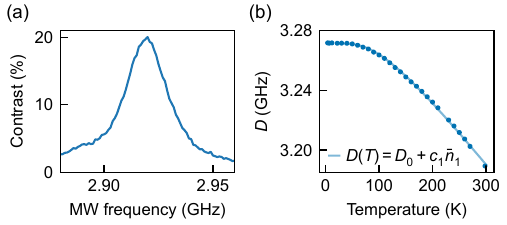}
\caption{(a) ODMR spectrum of the lowermost hyperfine transition with optimal power delivery shows a contrast of 20\%. (b) Zero-field splitting ($D$) as a function of temperature.}
\label{fig:zfs}
\end{figure}

We observe in Fig.~\ref{fig1}(e) that the ZFS at room temperature is smaller than that at 20~K. To measure ZFS across a range of temperatures, we use the PCB configuration optimized for microwave delivery at the lowermost hyperfine transition at 2.92~GHz, as described in Appendix \ref{si:plsetup}. Hence, we measure the ODMR frequency for this transition, $\nu$, as a function of temperature, and extract the ZFS ($D$) via
\begin{equation}
\nu = \frac{D}{2}+\frac{5A_{\parallel}}{4} + \frac{1}{2} \sqrt{\left(D+\frac{A_{\parallel}}{2}\right)^2 + 6A_{\perp}^2},
\end{equation}
where we assume that the fitted nuclear hyperfine constants have negligible temperature dependence. The values are plotted in Fig.~\ref{fig:zfs}(b). The decrease in ZFS with temperature results from thermal expansion of the diamond lattice and the increase of the atomic vibrational amplitude \cite{cambria2023physically}. The data show a trend similar to the temperature dependence of the ZFS of NV centers, which is modeled by \cite{cambria2023physically}
\begin{equation}
D(T) = D_0 + c_1 \bar{n}_1 + c_2 \bar{n}_2
\label{eq:zfs}
\end{equation}
where $D_0$ is the ZFS at 0~K, $c_{1(2)}$ describes how each phonon mode shifts the ZFS, and 
\begin{equation}
\bar{n}_{1(2)}=\frac{1}{e^{\epsilon_{1(2)}/k_B T}-1}
\label{eq:occupation}
\end{equation}
is the mean occupation number of the first (second) phonon mode with mode energy $\epsilon_{1(2)}$. Our dataset is well captured by considering one phonon mode, with fitted $D_0=3.2714\pm0.0002$~GHz, $\epsilon_1=25.64\pm 0.42$~meV, and $c_1=-137.3\pm3.7$~MHz. $\epsilon_1$ is similar to the first Orbach energy, $E_1=32.67$~meV, obtained from the temperature dependence of $T_1$ [Eq.~\eqref{T1rate}], indicating their common origin. Since our measurements only extend to room temperature ($k_B T \approx 25.7$~meV), phonon modes with energies $\epsilon \gg k_B T$ are weakly populated and contribute negligibly to the measured shift.

\section{Spin-dependent fluorescence}
\begin{figure}[b]
\includegraphics{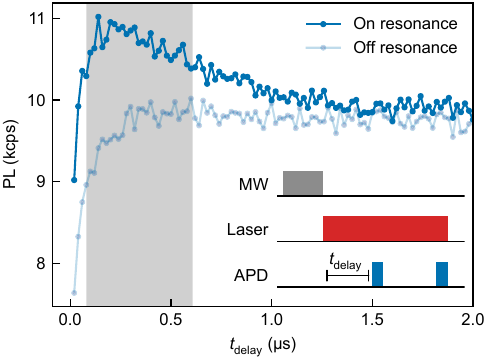}
\caption{PL as a function of the readout delay, $t_\mathrm{delay}$, with the microwave on resonance (dark blue) and off resonance (light blue). The shaded region marks the readout window used in the measurements. Inset: pulse sequence used for optimizing the readout window.}
\label{fig:spin-resolved}
\end{figure}

To determine the optimal detection window for spin dynamics measurements, we measure time-resolved spin-dependent PL using the pulse sequence shown in Fig.~\ref{fig:spin-resolved} with a detection window of 20~ns. After a resonant or non-resonant MW $\pi$-pulse, an optical pulse is turned on for spin readout and repolarization. A delay time before the integration window is swept to measure the repolarization rate. We observe spin-dependent PL up to $\sim$1~\us. Based on the response shown in Fig.~\ref{fig:spin-resolved}, we choose a repolarization time of 2~\us~and set the photon detection window to 600~ns. The response depends on the power density of the excitation spot. 

For CW ODMR and Rabi measurements, the contrast is normalized using $\Delta \text{PL}/\text{PL}=P_{\text{sig}}/P_{\text{ref}}-1$. For $T_1$ and $T_2$ measurements where the laser pulse duty cycle varies significantly with delay time, we use phase-inversion readout and normalize the data as $(P_{\text{sig}}-P_{\text{ref}})/(P_{\text{sig}}+P_{\text{ref}})$, as described in Ref.~\cite{zhang2020optically}.

\begin{figure}[t]
\includegraphics{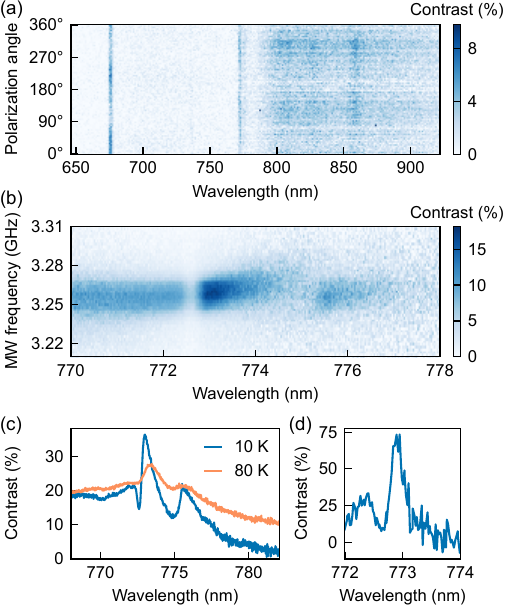}
    \caption{(a) Polarization dependence of spectrally resolved ODMR. Contribution from a single defect orientation is isolated by applying a magnetic field of 3~mT. Polarization angle of 26$^\circ$ in the lab frame is parallel to the defect axis. The integrated contrast for individual ZPLs is plotted in Fig.~\ref{fig2}(d) as a polar plot. (b) ODMR as a function of excitation wavelength shows a shift in the resonant frequency. This represents the ZFS distribution of the population probed by the narrow-linewidth resonant laser. (c) Wavelength dependence of ODMR in Sample A at 10~K and 80~K. MW frequency is fixed at 3.258~GHz. (d) Wavelength dependence of ODMR in Sample B achieves a maximum contrast of 73\% at 20~K. MW frequency is set to 2.92~GHz. }
    \label{fig:wvdep}
\end{figure}

\begin{figure}
\includegraphics{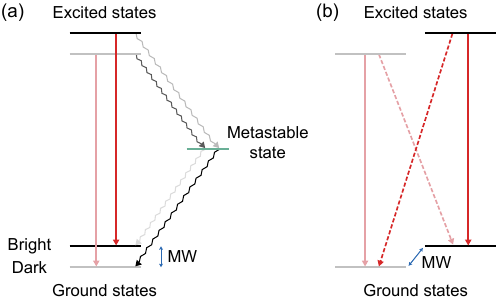}
    \caption{(a) Mechanism of ODMR and OSP via intersystem crossing to a metastable state. Straight arrows represent radiative transitions, with opacity indicating relative PL intensity. Wavy arrows represent nonradiative decay, with opacity indicating relative rates. (b) Mechanism of OSP via optical pumping. Solid and dashed lines represent spin-conserving and spin-flipping transitions, respectively. The ordering of bright and dark states is interchangeable.}
    \label{fig:mechanism}
\end{figure}

\section{Spectral decomposition and the temperature dependence of $T_2$}
\label{si:sd}
\begin{figure}[t!]
\includegraphics{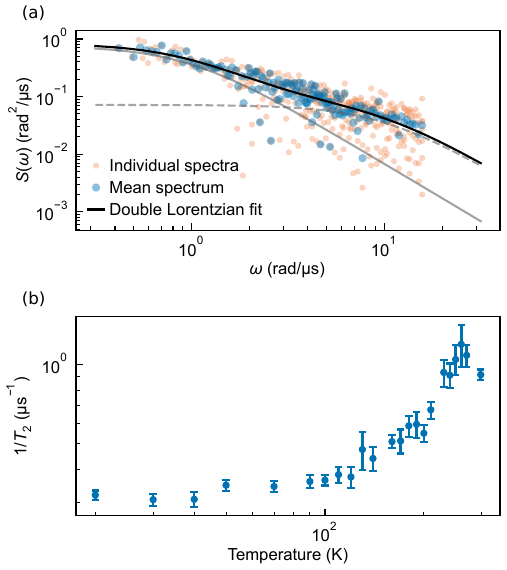}
\caption{(a) Noise spectrum from spectral decomposition. Orange and blue dots indicate
the individual traces computed for every $N_{\pi}$ and the mean spectrum, respectively. The black line is a fit to a double Lorentzian model [Eq.~\eqref{eq:double_lor}]. Solid and dashed gray lines are the individual Lorentzian
spectra. (b) Inverse of Hahn echo $T_2$ as a function of temperature.}
\label{fig:SD}
\end{figure}

Conventionally, the phase accumulated across a probed spin ensemble is assumed to follow a Gaussian distribution with a single variance. The noise spectrum is then extracted by \cite{BarGill2012}
\begin{equation}
S(\omega_0) = \frac{\pi \chi(t)}{t}, ~\omega_0 = \frac{\pi N_\pi}{t}
\label{eq:Sw_single}
\end{equation}
where $\chi(t)$ describes the coherence decay, $C(t)=e^{-\chi(t)}$. This approximation holds true for a mean-field model for bulk-doped defects \cite{BarGill2012, mukherjee2026defect}.  However, in our case, spins in a dense ensemble with implantation-induced disorder could experience different coupling strengths to the local noise bath, producing a broad, heavy-tailed distribution \cite{bauch2020decoherence}. Indeed, we find that the noise spectrum computed with Eq.~\eqref{eq:Sw_single} does not yield a self-consistent noise spectrum for our data: individual spectra extracted from $T_2$ decays under decoupling with different $N_{\pi}$ disagree systematically at the same frequency. Hence, we adopt the general formalism on decoherence dynamics developed in Ref.~\cite{Davis2023} that accounts for averaging over random bath configurations. For a three-dimensional dipolar bath, the ensemble coherence decays as 
\begin{equation}
\chi(t) = \Delta \sqrt{2\kappa(t)}
\label{eq:chi_ensemble}
\end{equation}
where $\Delta$ is the ensemble coupling strength that emerges from averaging the coherence over random positions of the bath spins, and 
\begin{equation}
\kappa(t) = \int_{-\infty}^{\infty} \frac{\dd\omega}{2\pi}\, S_1(\omega)\, \frac{F_{N_{\pi}}(\omega t)}{\omega^2}.
\label{eq:kappa}
\end{equation}
Here we introduce $S_1(\omega)$, the normalized spectrum of a single bath spin with 
$\int_{-\infty}^{\infty}\frac{\dd\omega}{2\pi}S_1(\omega)=1$, so that the total noise spectrum is $S(\omega) = \Delta^2 S_1(\omega)$.
Integrating the spectral filter function $F_{N_{\pi}}(\omega t)/\omega^2$ over its main peak centered at $\omega_0=\pi N_{\pi}/t$, Eq.~\eqref{eq:kappa} becomes
\begin{equation}
\kappa(t) \approx \frac{4t}{\pi^2} S_1(\omega_0).
\label{eq:kappa_integrated}
\end{equation}
This result differs from the ideal $\delta$-function approximation by a factor of $4/\pi$ but has no effect on the spectrum shape. Combining Eqs.~\eqref{eq:chi_ensemble} and \eqref{eq:kappa_integrated} yields
\begin{equation}
S(\omega_0) = \Delta^2 S_1(\omega_0)
 \approx \frac{\pi^2 (\ln{C})^2}{8t}.
\label{eq:Sw_ensemble}
\end{equation}
The noise spectrum computed from Eq.~\eqref{eq:Sw_ensemble} is plotted in Fig.~\ref{fig:SD}(a). We find the best fit to the noise spectrum to be a double Lorentzian model:
\begin{equation}
S(\omega) = \sum_{i=1,2} \frac{2\Delta_i^2 \tau_{c_i}}{1+(\omega \tau_{c_i})^2},
\label{eq:double_lor}
\end{equation}
where $\tau_{c_i}$ is the correlation time of the spin bath $i$ and $\Delta_i$ is the corresponding coupling strength. The fit yields  $\Delta_1=95.6\pm15.9$~kHz, $\Delta_2=94.5\pm6.9$~kHz, $\tau_{c_1}=1.05\pm0.10$~\us{}, and $\tau_{c_2}=102\pm44$~ns. The similar coupling strengths point to implantation-induced defects as candidates for the noise sources, such as gallium nuclear spins or unidentified gallium-related electron spins. We can eliminate such noise sources by studying the coherence properties of single centers in low-density implanted samples, which is left for future work. 

The CPMG scaling power $\gamma$ relates to the shape of the noise spectrum $S(\omega) \propto \omega ^{-\alpha}$ by $\gamma=\alpha/n$, where $n$ is the stretching factor in the $T_2$ decay \cite{mukherjee2026defect}. In our case, since $S(\omega)$ is proportional to $\chi^2$ instead of $\chi$ [Eq.~\eqref{eq:Sw_ensemble}], we have 
$\gamma = \frac{\alpha}{2n}$. From the fitted $S(\omega)$ [Fig.~\ref{fig:SD}(a)], we note that for smaller frequencies, the combination of two Lorentzian spectra is analogous to $1/f$-noise, i.e., $S(\omega) \propto \omega ^{-1}$. As $\omega$ is increased by applying larger $N_{\pi}$, the Lorentzian spectrum with the smaller $\tau_c$ gradually becomes the dominant contribution, so $S(\omega)$ approaches $\omega ^{-2}$. The fitted stretching factor $n$ is constant $\sim1$ for all $N_{\pi}$. Thus, we observe that $\gamma$ changes by a factor of $\sim2$ as larger $N_{\pi}$ is applied, giving rise to the characteristic slope change in Fig.~\ref{fig4}(d). 

Fig.~\ref{fig:SD}(b) shows $T_2$ as a function of temperature. $T_2$ decreases weakly with temperature, from $3.12\pm0.13$ \us{} at 20~K to $1.09\pm0.05$ \us{} at room temperature.

\section{Single-center measurements}
\label{si:single}

\begin{figure}[b!]
\includegraphics{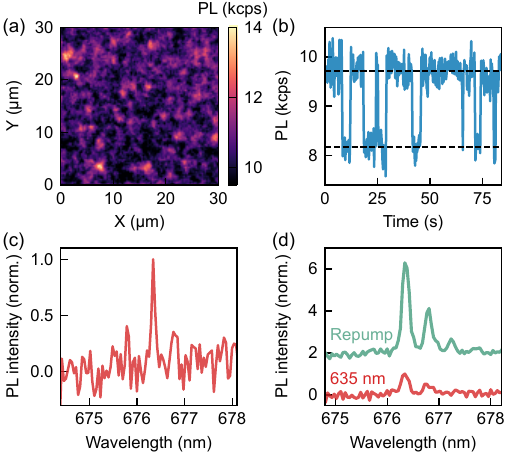}
\caption{(a) Confocal scan of Sample C, showing bright spots from gallium-related defects. The colorbar is adjusted to the background of 9.5~kcps. (b) The time trace of PL counts from a single center shows blinking from charge-state instability. (c) PL spectrum from an isolated emitter at 2~K. (d) PL spectra from a cluster of defects at 2~K show charge-state repumping with a 532~nm laser in addition to 635~nm excitation. An arbitrary offset is added for clarity.}
\label{fig:single}
\end{figure}
To search for single gallium-vacancy centers, we implant Sample C with $10^9$~ions/cm$^2$. Confocal scans show bright spots from gallium-related defects against a background emission of SiV$^-$ and other unidentified defects [Fig.~\ref{fig:single}(a)]. Since the counts from these bright spots are low relative to the background, we are unable to resolve antibunching in second-order correlation or $g^{(2)}(\tau)$ measurements to confirm single centers. However, charge-state switching of defects causes PL intermittency or blinking \cite{bradac2010observation} between two discrete levels [Fig.~\ref{fig:single}(b)], which is indicative of single emitters. The PL spectrum of such an emitter showing the 676~nm ZPL under off-resonant excitation is shown in Fig.~\ref{fig:single}(c). It exhibits narrow linewidths, limited by the spectrometer resolution ($\lesssim 50$~GHz). We also achieve charge-state repumping \cite{brevoord2024heralded} of a cluster of emitters using a continuous-wave 532~nm laser in addition to the 635~nm excitation. Fig.~\ref{fig:single}(d) demonstrates that the emission from the 676.4~nm ZPL is increased $\sim4\times$ under the repumping laser. Transitions to multiple vibronic levels in the ground state are also observable.
\bibliography{main}
\end{document}